\documentclass[a4paper,11pt]{article}
\usepackage{pos}
\usepackage{float}

\title{Non-perturbative three-nucleon simulation using chiral lattice EFT}

\author*[a]{Lukas Bovermann}
\author[a]{Evgeny Epelbaum}
\author[a]{Hermann Krebs}
\author[b]{Dean Lee}

\affiliation[a]{Institut f\"ur Theoretische Physik II, Fakult\"at f\"ur Physik und Astronomie, Ruhr-Universit\"at Bochum,\\
  Universit\"atsstra{\ss}e 150, D-44801 Bochum, Germany}

\affiliation[b]{Facility for Rare Isotope Beams and Department of Physics and Astronomy, Michigan State University,\\
  East Lansing, MI 48824, USA}

\emailAdd{lukas.bovermann@rub.de}
\emailAdd{evgeny.epelbaum@rub.de}
\emailAdd{hermann.krebs@rub.de}
\emailAdd{leed@frib.msu.edu}

\abstract{We study the three-nucleon system at next-to-next-to-next-to-leading order (N$^3$LO) in the framework of chiral effective field theory (EFT) on the lattice.
Our calculations do not rely on a perturbative treatment of subleading contributions to the nuclear forces.
For the two-nucleon potential, we apply the previously developed N$^3$LO lattice interaction.
For the leading contribution to the three-nucleon force, we determine the two low-energy constants (LECs) in the contact interactions by adjusting the ground state energy and half-life of triton, where the latter employs the nuclear axial current at N$^2$LO in chiral EFT.
Additionally, the ground state energy of helion and the charge radii of the two considered nuclei are computed.
No effect of the smearing regularization in the three-nucleon contact interaction is observed here.
We compare our results with recent lattice-EFT calculations that are based on a potential tuned to light and medium-mass nuclei using the wave-function-matching technique to circumvent the Monte-Carlo sign problem.}

\FullConference{
 The 11th International Workshop on Chiral Dynamics - CD2024\\
  26-30 August 2024\\
  Ruhr-Universit\"at Bochum, Germany
}

\begin{document}

\maketitle

\section{Introduction}
\label{sec:Intro}

\noindent
Nuclear lattice effective field theory (EFT) employs a discretized version of the chiral interaction between nucleons, pions and external currents (see Ref.~\cite{Lee:2008fa} for a review and Ref.~\cite{Lahde:2019npb} for a textbook).
Compared to lattice QCD, nuclear lattice EFT is advantageous for solving many-nucleon problems like bound states of medium-mass nuclei \cite{Elhatisari:2022zrb} and alpha-alpha scattering \cite{Elhatisari:2015iga}.
While applications of nuclear lattice EFT to many-nucleon systems require Monte Carlo simulations and perturbation theory beyond leading order to circumvent the fermion sign problem, few-nucleon systems can be treated non-perturbatively by solving the Schr\"odinger equation exactly.
We thus investigate in the following three-nucleon ground-state observables using the framework of chiral lattice EFT without relying on perturbation theory for the treatment of the higher-order nuclear forces.
In contrast to the recent lattice-EFT publications \cite{Elhatisari:2022zrb, Elhatisari:2024otn}, we do not consult data input from heavier nuclei.

\section{Chiral lattice interaction}
\label{sec:Interaction}

\begin{figure}[H]
\centering
\includegraphics{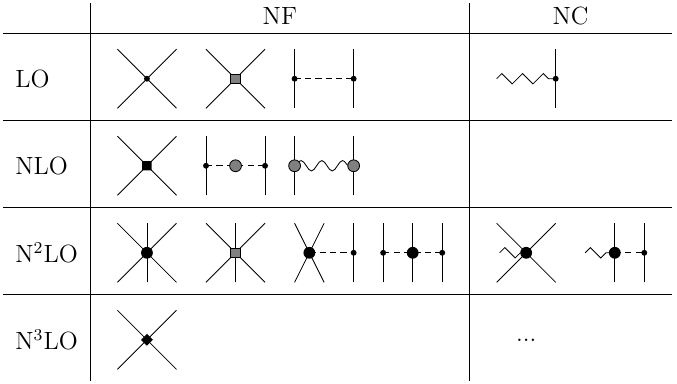}
\caption{Irreducible diagrams for nuclear force (NF) and axial nuclear current (NC) at (next-to-)$^n$leading order (N$^n$LO) of few-nucleon chiral lattice EFT.
The solid, dashed, wavy and zigzag lines denote nucleons, pions, photons and $\mathrm{W}^-$-bosons, respectively.
The black small circles, large circles, squares and diamonds correspond to vertices with the chiral dimension $0, 1, 2, 4$, respectively.
The gray circles indicate isospin-breaking Coulomb vertices and isospin-breaking pion propagators.
The gray squares indicate $\mathrm{SU}(4)$-symmetric vertices with additional local smearing compared to the (strictly speaking redundant) contact interactions in the diagrams next to the left.
We omit the nuclear axial current at N$^3$LO.
(The diagrams have been adopted from Refs.~\cite{Epelbaum:2008ga, Krebs:2016rqz}.)}
\label{fig:Interaction}
\end{figure}

\noindent
The nuclear force and nuclear axial current in few-nucleon chiral lattice EFT are given by the irreducible parts of the Feynman diagrams in Fig.~\ref{fig:Interaction}, which are explained in Refs.~\cite{Epelbaum:2008ga, Krebs:2016rqz}.
Contributions involving more than a single pion exchange between a pair of nucleons are neglected because they can be well represented by contact interactions for the large lattice spacing $a = 1.9733 \; \mathrm{fm}$ used here \cite{Li:2018ymw}.
We also neglect relativistic corrections proportional to the inverse nucleon mass \cite{Li:2018ymw} and axial-current terms vanishing in the beta-decay limit of a zero $\mathrm{W}^-$-boson momentum \cite{Baroni:2016xll}.

\noindent
Refs.~\cite{Epelbaum:2008ga, Krebs:2016rqz} list the expressions from the method of unitary transformation (MUT) corresponding to these diagrams, which must be discretized on the lattice.
The discretization of the kinetic-energy term in the Hamiltonian is $\mathcal{O}(a^4)$-improved as described in Ref.~\cite{Lee:2007jd}.
For the two-nucleon force, we apply the lattice implementation from Ref.~\cite{Li:2018ymw} consisting of
\begin{enumerate}
\item the 24 two-nucleon contact interactions fitted to the nucleon-nucleon phase shifts and mixing angles as well as to the deuteron ground state energy,

\item an additional two-nucleon contact interaction that possesses Wigner $\mathrm{SU}(4)$ symmetry and local smearing (important for many-nucleon systems \cite{Elhatisari:2016owd}),

\item the one-pion exchange potential with isospin-symmetry breaking due to the different pion masses and with a local Gaussian regulator,

\item the Coulomb interaction with the zero particle distance replaced by $a/2$,

\item the correction terms for the deuteron ground state energy in boosted frames that reduce the Galilean-invariance breaking mainly caused by the non-locally smeared contact interactions.
\end{enumerate}
For the three-nucleon force, we use the unsmeared interactions $V_{c_\mathrm{E}}^{(0)}$ (three-nucleon contact), $V_{c_\mathrm{D}}^{(0)}$ (single one-pion exchange) and $V_\mathrm{3N}^\mathrm{(TPE)}$ (double one-pion exchange) from Ref.~\cite{Elhatisari:2022zrb}, where only the pion exchange is regularized as described in item~3 additionally to the ultraviolet regulator introduced by the lattice spacing.
The low-energy constants (LECs) $c_\mathrm{E}$, $c_\mathrm{D}$ are fitted in Sec.~\ref{sec:Results}, and the LECs $c_1$, $c_3$, $c_4$ from $V_\mathrm{3N}^\mathrm{(TPE)}$ have been adjusted to pion-nucleon scattering data using Roy-Steiner equations in Ref.~\cite{Hoferichter:2015tha}.
We also introduce a locally smeared version $V_\mathrm{0, 3N}$ of $V_{c_\mathrm{E}}^{(0)}$ with the same smearing as in the two-nucleon $\mathrm{SU}(4)$-symmetric interaction \cite{Lu:2018bat}.
The nuclear axial current discretized from Ref.~\cite{Krebs:2016rqz} has the smearing properties, the pion-exchange regulator and the values of $c_\mathrm{D}$, $c_3$, $c_4$ analogous to the nuclear force.
Our remaining interaction parameters are chosen as in Ref.~\cite{Reinert:2017usi} and given by $m_\mathrm{N} = 938.92 \; \mathrm{MeV}$ (nucleon mass), $M_{\pi^0} = 134.98 \; \mathrm{MeV}$ (neutral-pion mass), $M_{\pi^\pm} = 139.57 \; \mathrm{MeV}$ (charged-pion mass), $F_\pi = 92.4 \; \mathrm{MeV}$ (pion decay constant), $g_\mathrm{A} = 1.29$ (effective nucleon axial-vector coupling constant), $\Lambda_\chi = 700 \; \mathrm{MeV}$ (convention for chiral-symmetry breaking scale \cite{Epelbaum:2002vt}) as well as $\alpha_\mathrm{EM} = 1/137.0360$ (fine-structure constant \cite{ParticleDataGroup:2022pth}).

\section{Three-nucleon results}
\label{sec:Results}

\noindent
Tab.~\ref{tab:Results} contains our results for the ground state energies and charge radii of triton ($^3 \mathrm{H}$) and helion ($^3 \mathrm{He}$) as well as the half-life of the beta decay between them.
The triton half-life was used to determine the value of the LEC $c_\mathrm{D}$, which has been followed by a fit of the triton ground state energy to fix either $c_\mathrm{E}$ or $C_\mathrm{0, 3N}$:
\begin{align}
c_\mathrm{D} = -0.0625 \pm 0.0625, \quad
c_\mathrm{E} = -0.39844 \pm 0.00781, \quad
C_\mathrm{0, 3N} = 0
\end{align}
or \footnote{Note that the LEC $C_\mathrm{0, 3N}$ is directly the coefficient of the smeared three-nucleon contact interaction $V_\mathrm{0, 3N}$, i.e.\ it has not been divided by a power of hard scales in order to make it naturally sized like $c_\mathrm{D}$ and $c_\mathrm{E}$.}
\begin{align}
c_\mathrm{D} = -0.0625 \pm 0.0625, \quad
C_\mathrm{0, 3N} = (-0.7246 \pm 0.0234) \; \mathrm{fm}^5, \quad
c_\mathrm{E} = 0.
\end{align}
For each fitting step, we bisected our LEC interval until we obtained only observable results closer to the experimental values than the truncation error bar (see next paragraph).
The half distance between the minimum and maximum value of the observable interval then served as a fitting-uncertainty estimation for the observable.
Propagating the $c_\mathrm{D}$-error through the triton-energy fit has been achieved by repeating the simulation for the extreme values of $c_\mathrm{D}$ and the central value of $c_\mathrm{E}$ or $C_\mathrm{0, 3N}$.

Another important source of uncertainty is caused by the truncation of the chiral expansion.
It is estimated here using the Epelbaum-Krebs-Mei{\ss}ner approach formulated in Ref.~\cite{Li:2018ymw} for an observable $X$ at N$^n$LO as
\begin{align}
\Delta X_{\mathrm{N}^n\mathrm{LO}} = \max \{ \mathcal{Q}^{n+2} \lvert X_\mathrm{LO} \rvert, \mathcal{Q}^n \lvert X_\mathrm{LO} - X_\mathrm{NLO} \rvert, \dots, \mathcal{Q} \lvert X_{\mathrm{N}^{n-1}\mathrm{LO}} - X_{\mathrm{N}^n\mathrm{LO}} \rvert \}.
\end{align}
Since all observables in this work are measured without a probe particle (ground state energies) or with a zero-momentum probe particle (charge radii and half-life), we choose the chiral expansion parameter momentum-independently as $\mathcal{Q} = M_\pi^\mathrm{eff} / \Lambda_\mathrm{b}$ with the effective pion mass $M_\pi^\mathrm{eff} = 200 \; \mathrm{MeV}$ and the breakdown scale $\Lambda_\mathrm{b} = 600 \; \mathrm{MeV}$ (cf.\ Ref.~\cite{Epelbaum:2019wvf}).
With the two-nucleon contact LECs available only at N$^3$LO in Ref.~\cite{Li:2018ymw}, we can just produce NLO results by omitting the three-nucleon force and the two-nucleon axial current; we thus set $X_\mathrm{LO} = X_\mathrm{NLO}$ and $X_\mathrm{N^2LO} = X_\mathrm{N^3LO}$.
Given the small amount of results at low orders and the neglected N$^3$LO contribution to the nuclear axial current, our estimations of the truncation uncertainty should be taken with care.
Finally, the fitting errors of the two LECs and the truncation error are added in quadrature to yield the total error listed in Tab.~\ref{tab:Results}.

We employ a periodic cubic lattice with length $L$ and spacing $a = 1.9733 \; \mathrm{fm}$; details of the observable computations can be found in Secs.~\ref{sec:Results_EB}--\ref{sec:Results_t12}.

\begin{table}[H]
\centering
\begin{tabular}{l|l|l|l|l|l}
& $E_\mathrm{3H} \; [\mathrm{MeV}]$ & $E_\mathrm{3He} \; [\mathrm{MeV}]$ & $R_\mathrm{ch, 3H} \; [\mathrm{fm}]$ & $R_\mathrm{ch, 3He} \; [\mathrm{fm}]$ & $t_{1/2} \; \mathrm{[yr]}$ \\
\hline
NLO & $-7.56(26)$ & $-6.86(25)$ & $1.79(6)$ & $2.04(6)$ & $11.7(7)$ \\
$c_\mathrm{D}+c_\mathrm{E}$-fit & $-8.48(7)$ & $-7.73(7)$ & $1.695(10)$ & $1.914(14)$ & $12.31(10)$ \\
$c_\mathrm{D}+C_\mathrm{0,3N}$-fit & $-8.51(8)$ & $-7.77(7)$ & $1.702(10)$ & $1.920(13)$ & $12.30(10)$ \\
\hline
Ref.~\cite{Elhatisari:2022zrb} * & $-8.35(18)$ & $-7.64(17)$ & $1.713(9)$ & $1.901(28)$ & \\
Ref.~\cite{Elhatisari:2024otn} ** & $-8.33(2)$ & $-7.62(2)$ & & & $12.13(8)$ \\
\hline
experiment & $-8.481795(2)$ & $-7.718040(2)$ & $1.7591(363)$ & $1.9661(30)$ & $12.32(3)$
\end{tabular}
\caption{Triton ground state energy $E_\mathrm{3H}$ (fitted using LEC $c_\mathrm{E}$ or $C_\mathrm{0, 3N}$), helion ground state energy $E_\mathrm{3He}$, triton charge radius $R_\mathrm{ch, 3H}$, helion charge radius $R_\mathrm{ch, 3He}$ and triton half-life $t_{1/2}$ (fitted using LEC $c_\mathrm{D}$) in comparison with NLO predictions, other perturbative \cite{Elhatisari:2022zrb} and non-perturbative \cite{Elhatisari:2024otn} lattice-EFT results based on a potential tuned to light and medium-mass nuclei as well as experimental data from Refs.~\cite{Wang:2017, Angeli:2013epw, Simpson:1987zz}. \\
* We have added the computational error and the chiral-interaction uncertainty from Ref.~\cite{Elhatisari:2022zrb} in quadrature. The additional infinite projection time extrapolation and Monte Carlo error may explain why the total uncertainty is mostly larger than in our work. \\
** This reference does not take the chiral truncation uncertainty into account, which could be the reason for the experimental data lying outside the theoretical error bars.}
\label{tab:Results}
\end{table}

\subsection{Ground state energies}
\label{sec:Results_EB}

\noindent
The ground state energy $E$ of a nucleus is obtained by calculating the lowest eigenenergy of the Hamiltonian matrix using the self-implemented Lanczos algorithm described in Ref.~\cite{Stellin:2018fkj}.
After performing this computation for the lattice lengths $L \in \{ 4a, 5a, \dots, 9a \}$, one can extrapolate $E$ to infinite volume using the $L$-dependence
\begin{align}
E(L \to \infty) = E_\infty + E_0 L^{-3/2} \exp( -L / L_0 )
\end{align}
of a three-particle ground state energy derived in Ref.~\cite{Meissner:2014dea}.
The truncation uncertainty of the individual data points is propagated to infinite lattice length via statistical sampling.

\subsection{Charge radii}
\label{sec:Results_Rch}

\noindent
The mean square charge radius of a nucleus is defined in the absence of the two- and more-nucleon contributions to the charge density operator according to Refs.~\cite{Hoppe:2019uyw, Simonis:2017dny, Foldy:1949wa, Friar:1997js} as
\begin{align}
R_\mathrm{ch}^2 = \langle \psi \rvert R^2_\mathrm{p} \lvert \psi \rangle + \left( R_\mathrm{ch, p}^2 + \frac{N}{Z} R_\mathrm{ch, n}^2 \right) + \frac{3}{4 m_\mathrm{p}^2},
\end{align}
i.e.\ it is the sum of
\begin{enumerate}
\item the squared point-proton radius $\langle \psi \rvert R^2_\mathrm{p} \lvert \psi \rangle$ obtained as an expectation value using the nuclear wave function $\psi$, which is the Hamiltonian eigenvector corresponding to the lowest eigenenergy computed for $L = 9 a$ by the Lanczos algorithm,

\item the corrections for the non-zero charge radii of the $Z$ protons ($R_\mathrm{ch, p}^2 = (0.7071 \pm 0.0007) \; \mathrm{fm}^2$ \cite{ParticleDataGroup:2022pth}) and the $N$ neutrons ($R_\mathrm{ch, n}^2 = (-0.1155 \pm 0.0017) \; \mathrm{fm}^2$ \cite{ParticleDataGroup:2022pth}),

\item the relativistic Darwin-Foldy correction with the proton mass $m_\mathrm{p} = 938.27208816(29) \; \mathrm{MeV}$ from Ref.~\cite{ParticleDataGroup:2022pth}.
\end{enumerate}
We do not consider the spin-orbit correction because it vanishes in the nuclear ground state with zero orbital angular momenta of all nucleons. \cite{Ong:2010gf}
The truncation uncertainty of the point-proton radius is Gaussian-propagated to the charge radius.

\subsection{Half-life}
\label{sec:Results_t12}

\noindent
According to Refs.~\cite{Baroni:2016xll, Schiavilla:1998je}, the half-life of the triton beta decay can be computed via
\begin{align}
t_{1/2} = \frac{1}{(1 + \delta_\mathrm{R}) f_\mathrm{V}} \frac{K / G_\mathrm{V}^2}{\langle \mathbf{F} \rangle^2 + f_\mathrm{A} / f_\mathrm{V} g_\mathrm{A}^2 \langle \mathbf{GT} \rangle^2}
\end{align}
with
\begin{enumerate}
\item the Fermi matrix element $\langle \mathbf{F} \rangle = \sum_{n = 1}^3 \langle \psi_\mathrm{3He} \rvert \rvert \tau_{n,+} \lvert \lvert \psi_\mathrm{3H} \rangle$ for the same nuclear wave functions as in Sec.~\ref{sec:Results_Rch}, where the isospin-raising operator $\tau_{n,+}$ for the $n$-th nucleon must be reduced in the sense of the Wigner-Eckart theorem \cite{Schiavilla:1998je, Raman:1978qta},

\item the Gamow-Teller matrix element $\langle \mathbf{GT} \rangle = -\frac{2}{g_\mathrm{A}} \langle \psi_\mathrm{3He} \rvert \rvert \vec{A}^+ (q^2 \to 0) \lvert \lvert \psi_\mathrm{3H} \rangle$, where the nuclear axial current $\vec{A}^+$ \cite{Krebs:2016rqz} at zero squared $\mathrm{W}^-$-boson momentum transfer $q^2$ to the nucleus is a raising operator in isospin space and a three-vector in coordinate space \cite{Schiavilla:1998je, Raman:1978qta},

\item the numerical constants $\delta_\mathrm{R} = 1.90 \; \%$ (outer radiative correction \cite{Raman:1978qta}), $f_\mathrm{A} = 2.8505 \times 10^{-6}$ and $f_\mathrm{V} = 2.8355 \times 10^{-6}$ (Fermi functions \cite{Simpson:1987zz}), $K / G_\mathrm{V}^2 = (6144.5 \pm 1.9) \; \mathrm{s}$ \cite{Baroni:2016xll, Hardy:2014qxa}.
\end{enumerate}
The truncation uncertainties of $\lvert \langle \mathbf{F} \rangle \rvert$ and $\lvert \langle \mathbf{GT} \rangle \rvert$ are Gaussian-propagated to the half-life.

\section{Summary and outlook}
\label{sec:SummOut}

\noindent
The exploratory non-perturbative three-nucleon simulation performed here in the framework of nuclear lattice EFT at N$^3$LO has lead to reasonable results (cf.\ Tab.~\ref{tab:Results}) after fitting the triton ground state energy and triton half-life.
The addition of the three-nucleon force at N$^2$LO also cured the deviation of the helion ground state energy still present at NLO.
The predicted charge radii are a few percent smaller than their experimental values, which might be caused by neglecting the exchange contributions to the charge density operator in our analysis.
We have found no significant difference between fitting the three-nucleon contact interaction with and without local smearing; the effect of the locality observed in Ref.~\cite{Elhatisari:2016owd} therefore likely only occurs in systems with more than three nucleons.

Including bound state energies of heavier nuclei into the LEC fits as in the perturbative Monte Carlo computation \cite{Elhatisari:2022zrb} (cf.\ Tab.~\ref{tab:Results}) does not seem to improve the three-nucleon charge radii significantly.
It also appears to be insufficient for an accurate description of the triton half-life if the nuclear axial current is only considered at leading order and the truncation uncertainty is neglected as in the non-perturbative study \cite{Elhatisari:2024otn} (see the same table).

The interaction constructed here could now be applied e.g.\ in nucleon-deuteron scattering via the adiabatic projection method \cite{Pine:2013zja}.

\acknowledgments{
\noindent
We are grateful to Serdar Elhatisari, Yuanzhuo Ma, Ulf-G. Mei{\ss}ner and other members of the Nuclear Lattice Effective Field Theory Collaboration for helpful discussions.
We thank Serdar Elhatisari for sharing his code for the lattice-EFT Hamiltonian.
This work was supported by ERC AdG NuclearTheory (Grant No.\ 885150), by DFG and NSFC through funds provided to the Sino-German CRC 110 ``Symmetries and the Emergence of Structure in QCD'' (NSFC Grant No.\ 11621131001, DFG Project-ID 196253076 - TRR 110), by the MKW NRW under the funding code NW21-024-A, by the EU Horizon 2020 research and innovation programme (STRONG-2020, grant agreement No.\ 824093), and by the US Department of Energy grants DE-SC0013365, DE-SC0023175, and DE-SC0024586.
}

\bibliographystyle{JHEP}
\bibliography{literature_proceedings.bib}

\providecommand{\href}[2]{#2}\begingroup\raggedright\begin{thebibliography}{10}

\bibitem{Lee:2008fa}
D.~Lee, \emph{{Lattice simulations for few- and many-body systems}},
  \href{https://doi.org/10.1016/j.ppnp.2008.12.001}{\emph{Prog. Part. Nucl.
  Phys.} {\bfseries 63} (2009) 117}
  [\href{https://arxiv.org/abs/0804.3501}{{\ttfamily 0804.3501}}].

\bibitem{Lahde:2019npb}
T.A.~L\"ahde and U.-G.~Mei\ss{}ner, \emph{{Nuclear Lattice Effective Field
  Theory}: {An introduction}}, vol.~957, Springer (2019),
  \href{https://doi.org/10.1007/978-3-030-14189-9}{10.1007/978-3-030-14189-9}.

\bibitem{Elhatisari:2022zrb}
S.~Elhatisari et~al., \emph{{Wavefunction matching for solving quantum
  many-body problems}},
  \href{https://doi.org/10.1038/s41586-024-07422-z}{\emph{Nature} {\bfseries
  630} (2024) 59} [\href{https://arxiv.org/abs/2210.17488}{{\ttfamily
  2210.17488}}].

\bibitem{Elhatisari:2015iga}
S.~Elhatisari, D.~Lee, G.~Rupak, E.~Epelbaum, H.~Krebs, T.A.~L\"ahde et~al.,
  \emph{{Ab initio alpha-alpha scattering}},
  \href{https://doi.org/10.1038/nature16067}{\emph{Nature} {\bfseries 528}
  (2015) 111} [\href{https://arxiv.org/abs/1506.03513}{{\ttfamily
  1506.03513}}].

\bibitem{Elhatisari:2024otn}
S.~Elhatisari, F.~Hildenbrand and U.-G.~Mei\ss{}ner, \emph{{The triton lifetime
  from nuclear lattice effective field theory}},
  \href{https://doi.org/10.1016/j.physletb.2024.139086}{\emph{Phys. Lett. B}
  {\bfseries 859} (2024) 139086}
  [\href{https://arxiv.org/abs/2408.06670}{{\ttfamily 2408.06670}}].

\bibitem{Epelbaum:2008ga}
E.~Epelbaum, H.-W.~Hammer and U.-G.~Mei{\ss}ner, \emph{{Modern Theory of
  Nuclear Forces}},
  \href{https://doi.org/10.1103/RevModPhys.81.1773}{\emph{Rev. Mod. Phys.}
  {\bfseries 81} (2009) 1773}
  [\href{https://arxiv.org/abs/0811.1338}{{\ttfamily 0811.1338}}].

\bibitem{Krebs:2016rqz}
H.~Krebs, E.~Epelbaum and U.-G.~Mei\ss{}ner, \emph{{Nuclear axial current
  operators to fourth order in chiral effective field theory}},
  \href{https://doi.org/10.1016/j.aop.2017.01.021}{\emph{Annals Phys.}
  {\bfseries 378} (2017) 317}
  [\href{https://arxiv.org/abs/1610.03569}{{\ttfamily 1610.03569}}].

\bibitem{Li:2018ymw}
N.~Li, S.~Elhatisari, E.~Epelbaum, D.~Lee, B.-N.~Lu and U.-G.~Mei\ss{}ner,
  \emph{{Neutron-proton scattering with lattice chiral effective field theory
  at next-to-next-to-next-to-leading order}},
  \href{https://doi.org/10.1103/PhysRevC.98.044002}{\emph{Phys. Rev. C}
  {\bfseries 98} (2018) 044002}
  [\href{https://arxiv.org/abs/1806.07994}{{\ttfamily 1806.07994}}].

\bibitem{Baroni:2016xll}
A.~Baroni, L.~Girlanda, A.~Kievsky, L.E.~Marcucci, R.~Schiavilla and
  M.~Viviani, \emph{{Tritium $\beta$-decay in chiral effective field theory}},
  \href{https://doi.org/10.1103/PhysRevC.94.024003}{\emph{Phys. Rev. C}
  {\bfseries 94} (2016) 024003}
  [\href{https://arxiv.org/abs/1605.01620}{{\ttfamily 1605.01620}}].

\bibitem{Lee:2007jd}
D.~Lee and R.~Thomson, \emph{{Temperature-dependent errors in nuclear lattice
  simulations}}, \href{https://doi.org/10.1103/PhysRevC.75.064003}{\emph{Phys.
  Rev. C} {\bfseries 75} (2007) 064003}
  [\href{https://arxiv.org/abs/nucl-th/0701048}{{\ttfamily nucl-th/0701048}}].

\bibitem{Elhatisari:2016owd}
S.~Elhatisari et~al., \emph{{Nuclear binding near a quantum phase transition}},
  \href{https://doi.org/10.1103/PhysRevLett.117.132501}{\emph{Phys. Rev. Lett.}
  {\bfseries 117} (2016) 132501}
  [\href{https://arxiv.org/abs/1602.04539}{{\ttfamily 1602.04539}}].

\bibitem{Hoferichter:2015tha}
M.~Hoferichter, J.~Ruiz~de Elvira, B.~Kubis and U.-G.~Mei\ss{}ner,
  \emph{{Matching pion-nucleon Roy-Steiner equations to chiral perturbation
  theory}}, \href{https://doi.org/10.1103/PhysRevLett.115.192301}{\emph{Phys.
  Rev. Lett.} {\bfseries 115} (2015) 192301}
  [\href{https://arxiv.org/abs/1507.07552}{{\ttfamily 1507.07552}}].

\bibitem{Lu:2018bat}
B.-N.~Lu, N.~Li, S.~Elhatisari, D.~Lee, E.~Epelbaum and U.-G.~Mei\ss{}ner,
  \emph{{Essential elements for nuclear binding}},
  \href{https://doi.org/10.1016/j.physletb.2019.134863}{\emph{Phys. Lett. B}
  {\bfseries 797} (2019) 134863}
  [\href{https://arxiv.org/abs/1812.10928}{{\ttfamily 1812.10928}}].

\bibitem{Reinert:2017usi}
P.~Reinert, H.~Krebs and E.~Epelbaum, \emph{{Semilocal momentum-space
  regularized chiral two-nucleon potentials up to fifth order}},
  \href{https://doi.org/10.1140/epja/i2018-12516-4}{\emph{Eur. Phys. J. A}
  {\bfseries 54} (2018) 86} [\href{https://arxiv.org/abs/1711.08821}{{\ttfamily
  1711.08821}}].

\bibitem{Epelbaum:2002vt}
E.~Epelbaum, A.~Nogga, W.~Gl{\"o}ckle, H.~Kamada, U.-G.~Mei{\ss}ner and
  H.~Wita{\l}a, \emph{{Three nucleon forces from chiral effective field
  theory}}, \href{https://doi.org/10.1103/PhysRevC.66.064001}{\emph{Phys. Rev.
  C} {\bfseries 66} (2002) 064001}
  [\href{https://arxiv.org/abs/nucl-th/0208023}{{\ttfamily nucl-th/0208023}}].

\bibitem{ParticleDataGroup:2022pth}
{\scshape Particle Data Group} collaboration, \emph{{Review of Particle
  Physics}}, \href{https://doi.org/10.1093/ptep/ptac097}{\emph{PTEP} {\bfseries
  2022} (2022) 083C01}.

\bibitem{Epelbaum:2019wvf}
E.~Epelbaum, \emph{{High-precision nuclear forces : Where do we stand?}},
  \href{https://doi.org/10.22323/1.317.0006}{\emph{PoS} {\bfseries CD2018}
  (2019) 006}.

\bibitem{Wang:2017}
M.~Wang, G.~Audi, F.G.~Kondev, W.J.~Huang, S.~Naimi and X.~Xu, \emph{{The
  AME2016 atomic mass evaluation (II). Tables, graphs and references}},
  \href{https://doi.org/10.1088/1674-1137/41/3/030003}{\emph{Chin. Phys. C}
  {\bfseries 41} (2017) 030003}.

\bibitem{Angeli:2013epw}
I.~Angeli and K.P.~Marinova, \emph{{Table of experimental nuclear ground state
  charge radii: An update}},
  \href{https://doi.org/10.1016/j.adt.2011.12.006}{\emph{Atom. Data Nucl. Data
  Tabl.} {\bfseries 99} (2013) 69}.

\bibitem{Simpson:1987zz}
J.J.~Simpson, \emph{{Half-life of tritium and the Gamow-Teller transition
  rate}}, \href{https://doi.org/10.1103/PhysRevC.35.752}{\emph{Phys. Rev. C}
  {\bfseries 35} (1987) 752}.

\bibitem{Stellin:2018fkj}
G.~Stellin, S.~Elhatisari and U.-G.~Mei\ss{}ner, \emph{{Breaking and
  restoration of rotational symmetry in the low-energy spectrum of light
  alpha-conjugate nuclei on the lattice I: $^{8}\mathrm{Be}$ and
  $^{12}\mathrm{C}$}},
  \href{https://doi.org/10.1140/epja/i2018-12671-6}{\emph{Eur. Phys. J. A}
  {\bfseries 54} (2018) 232}
  [\href{https://arxiv.org/abs/1809.06109}{{\ttfamily 1809.06109}}].

\bibitem{Meissner:2014dea}
U.-G.~Mei\ss{}ner, G.~R\'\i{}os and A.~Rusetsky, \emph{{Spectrum of three-body
  bound states in a finite volume}},
  \href{https://doi.org/10.1103/PhysRevLett.117.069902}{\emph{Phys. Rev. Lett.}
  {\bfseries 114} (2015) 091602}
  [\href{https://arxiv.org/abs/1412.4969}{{\ttfamily 1412.4969}}].

\bibitem{Hoppe:2019uyw}
J.~Hoppe, C.~Drischler, K.~Hebeler, A.~Schwenk and J.~Simonis, \emph{{Probing
  chiral interactions up to next-to-next-to-next-to-leading order in
  medium-mass nuclei}},
  \href{https://doi.org/10.1103/PhysRevC.100.024318}{\emph{Phys. Rev. C}
  {\bfseries 100} (2019) 024318}
  [\href{https://arxiv.org/abs/1904.12611}{{\ttfamily 1904.12611}}].

\bibitem{Simonis:2017dny}
J.~Simonis, S.R.~Stroberg, K.~Hebeler, J.D.~Holt and A.~Schwenk,
  \emph{{Saturation with chiral interactions and consequences for finite
  nuclei}}, \href{https://doi.org/10.1103/PhysRevC.96.014303}{\emph{Phys. Rev.
  C} {\bfseries 96} (2017) 014303}
  [\href{https://arxiv.org/abs/1704.02915}{{\ttfamily 1704.02915}}].

\bibitem{Foldy:1949wa}
L.L.~Foldy and S.A.~Wouthuysen, \emph{{On the Dirac theory of spin 1/2 particle
  and its nonrelativistic limit}},
  \href{https://doi.org/10.1103/PhysRev.78.29}{\emph{Phys. Rev.} {\bfseries 78}
  (1950) 29}.

\bibitem{Friar:1997js}
J.L.~Friar, J.~Martorell and D.W.L.~Sprung, \emph{{Nuclear sizes and the
  isotope shift}}, \href{https://doi.org/10.1103/PhysRevA.56.4579}{\emph{Phys.
  Rev. A} {\bfseries 56} (1997) 4579}
  [\href{https://arxiv.org/abs/nucl-th/9707016}{{\ttfamily nucl-th/9707016}}].

\bibitem{Ong:2010gf}
A.~Ong, J.C.~Berengut and V.V.~Flambaum, \emph{{The Effect of spin-orbit
  nuclear charge density corrections due to the anomalous magnetic moment on
  halonuclei}}, \href{https://doi.org/10.1103/PhysRevC.82.014320}{\emph{Phys.
  Rev. C} {\bfseries 82} (2010) 014320}
  [\href{https://arxiv.org/abs/1006.5508}{{\ttfamily 1006.5508}}].

\bibitem{Schiavilla:1998je}
R.~Schiavilla et~al., \emph{{Weak capture of protons by protons}},
  \href{https://doi.org/10.1103/PhysRevC.58.1263}{\emph{Phys. Rev. C}
  {\bfseries 58} (1998) 1263}
  [\href{https://arxiv.org/abs/nucl-th/9808010}{{\ttfamily nucl-th/9808010}}].

\bibitem{Raman:1978qta}
S.~Raman, C.A.~Houser, T.A.~Walkiewicz and I.S.~Towner, \emph{{Mixed Fermi and
  Gamow-Teller \ensuremath{\beta}-transitions and isoscalar magnetic moments}},
  \href{https://doi.org/10.1016/0092-640X(78)90008-6}{\emph{Atom. Data Nucl.
  Data Tabl.} {\bfseries 21} (1978) 567}.

\bibitem{Hardy:2014qxa}
J.C.~Hardy and I.S.~Towner, \emph{{Superallowed $0^+\to 0^+$ nuclear
  \ensuremath{\beta} decays: 2014 critical survey, with precise results for
  $V_{ud}$ and CKM unitarity}},
  \href{https://doi.org/10.1103/PhysRevC.91.025501}{\emph{Phys. Rev. C}
  {\bfseries 91} (2015) 025501}
  [\href{https://arxiv.org/abs/1411.5987}{{\ttfamily 1411.5987}}].

\bibitem{Pine:2013zja}
M.~Pine, D.~Lee and G.~Rupak, \emph{{Adiabatic projection method for scattering
  and reactions on the lattice}},
  \href{https://doi.org/10.1140/epja/i2013-13151-3}{\emph{Eur. Phys. J. A}
  {\bfseries 49} (2013) 151} [\href{https://arxiv.org/abs/1309.2616}{{\ttfamily
  1309.2616}}].

\end{thebibliography}\endgroup

\end{document}